\documentclass[]{emulateapj}
\shorttitle{Major mergers and the rise of early types}
\shortauthors{L\'opez-Sanjuan et al.}

\begin{document}

\title{The minor role of gas-rich major mergers in the rise of\\
intermediate-mass early types at $z \leq 1$}

\author{Carlos L\'opez-Sanjuan\altaffilmark{1,2,3}, Marc Balcells\altaffilmark{1,2,4},
Pablo G. P\'erez-Gonz\'alez\altaffilmark{5}, Guillermo Barro\altaffilmark{5},
C\'esar Enrique Garc\'{i}a-Dab\'o\altaffilmark{1,6}, Jes\'us
Gallego\altaffilmark{5}, Jaime Zamorano\altaffilmark{5}}

\altaffiltext{1}{Instituto de Astrof\'{\i}sica de Canarias, Calle V\'{\i}a
L\'actea s/n, E-38205 La Laguna, Tenerife, Spain}
\altaffiltext{2}{Departamento de Astrof\'\i sica, Universidad de La Laguna, 
E-38200 La Laguna, Tenerife, Spain}
\altaffiltext{3}{Laboratoire d'Astrophysique de Marseille, P\^ole de l'Etoile Site de Ch\^ateau-Gombert 38, rue Fr\'ed\'eric Joliot-Curie, F-13388 Marseille, France}
\altaffiltext{4}{Isaac Newton Group of Telescopes, Aptdo.~Correos 321, E-38700 Santa Cruz de La Palma, Tenerife, Spain}
\altaffiltext{5}{Departamento de Astrof\'{\i}sica y Ciencias de la Atm\'osfera,
Facultad de C.C. F\'{\i}sicas, Universidad Complutense de Madrid, E-28040
Madrid, Spain}
\altaffiltext{6}{FRACTAL SLNE, C/ Tulip\'an 2, portal 13, 1 A, E-28231 Las Rozas, Madrid, Spain} 

\email{clsj@iac.es}

\begin{abstract}
We study the evolution of galaxy structure since $z \sim 1$ to the present. 
From a Great Observatories Origins Deel Survey South (GOODS-S) multi-band catalog we define (blue) luminosity- and mass-weighted samples, limited by $M_B \leq -20$ and $M_{\star} \geq 10^{10}\ M_{\odot}$, comprising 1122 and 987 galaxies, respectively.
We extract early-type (ET; E/S0/Sa) and late-type (LT; Sb-Irr) subsamples by their position in the concentration--asymmetry plane, in which galaxies exhibit a clear bimodality. We find that the ET fraction, $f_{\rm ET}$, rises with cosmic time, with a corresponding decrease in the LT fraction, $f_{\rm LT}$, in both luminosity- and mass-selected samples. However, the evolution of the comoving number density is very
different: the decrease in the total number density of $M_B \leq -20$ galaxies
since $z = 1$ is due to the decrease in the LT population, which accounts
for $\sim75$\% of the total star formation rate in the range under study, while 
the increase in the total number density of $M_{\star} \geq 10^{10}\ M_{\odot}$
galaxies in the same redshift range is due to the evolution of ETs. This suggests
that we need a structural transformation between LT galaxies that form
stars actively and ET galaxies in which the stellar mass is located. Comparing the observed evolution with the gas-rich major merger rate in
GOODS-S, we infer that only $\sim$20\% of the new ET galaxies with
$M_{\star} \geq 10^{10}\ M_{\odot}$ appeared since $z \sim 1$ can be explained by
this kind of mergers, suggesting that minor mergers and secular processes may be the driving mechanisms of the structural evolution of intermediate-mass ($M_{\star} \sim 4 \times 10^{10}\
M_{\odot}$) galaxies since $z \sim 1$.
\end{abstract}


\section{introduction}
Local galaxies present two main populations in the color-magnitude diagram:
the red sequence, formed primarily by old, spheroidal quiescent galaxies, and
the blue cloud, formed primarily by spiral star-forming galaxies
\citep{strateva01, baldry04}. It is now well established that such bimodality is present at higher
redshifts (\citealt{bell04}, up to $z \sim 1$; \citealt{cassata08,ilbert10}, up
to $z \sim 2$; \citealt{kriek08}, at $z \sim 2.3$), and appears to be strongly linked to mass: more massive
galaxies were the first to finish forming their stars and populating the red sequence. This mass dependence has been dubbed downsizing \citep{cowie96}: massive galaxies having
experienced most of their star formation at early times and being passive by $z
\sim 1$, and many among the less massive galaxies experience extended
star-formation histories \citep[see][and references
therein]{bundy06,scarlata07ee,pgon08}.

These results are not immediately expected from the popular hierarchical $\Lambda$-CDM models,
in which the more massive dark matter halos are the final stage
of successive mergers of smaller halos. However, whether downsizing poses an important problem for $\Lambda$-CDM depends on the correct understanding of the baryonic physics, for which models are making recent progress \citep[see][and references
therein]{bower06,delucia07,stewart09,hopkins09bulges}, as well as on proper accounting of the dependency of the halo merger history with environment. Because of that, the
role of galaxy mergers in the buildup of the red sequence, and their impact on the evolution of galaxy properties, i.e., color, mass, or structure, remains an important open question.

In addition, the redshift evolution of the mass function suggests that the red sequence grows
because star formation is quenched in the blue cloud \citep{bell07,ruhland09}.
Because $\sim$80\% of red sequence galaxies are morphological early types (ETs) at $z \lesssim 1$
\citep[E/S0/Sa,][]{strateva01,lotz08ff}, and because the star
formation is located in spiral galaxies \citep{bell05,jogee09}, the
blue-to-red transition may be accompanied by a late- to ET
transformation.

In this paper we study the role of gas-rich major mergers, selected by
morphological criteria, in the structural evolution of galaxies since $z \sim
1$. In a previous paper, \citet[][hereafter L09]{clsj09ffgoods}, we study the major merger rate evolution in GOODS-S, finding that only $\sim 10$\% of $z = 0$ galaxies with $M_{\star} \geq 10^{10}\ M_{\odot}$ have undergone a gas-rich major merger since $z \sim 1$. Our goal is, first to quantify the structural evolution, and, second, to compare it against the merger rate.  

The paper is organized as follows: in Section~\ref{data} we summarize the GOODS-S
data set used, and in Section~\ref{cabimo} we develop the
methodology to determine the fractions of ET and late-type (LT) galaxies versus redshift. Then, in Section~\ref{results} we summarize the obtained
early- and LT fractions and comoving number densities, and their
evolution with $z$, while in Section~\ref{rolewet} we study the role of
gas-rich major mergers in the observed ET evolution. Finally, we present our conclusions in Section~\ref{conclusion}. We use $H_0 = 70\ {\rm km\
s^{-1}\ Mpc^{-1}}$, $\Omega_{M} = 0.3$, and $\Omega_{\Lambda} = 0.7$ throughout.
All magnitudes are Vega unless noted otherwise.

\section{GOODS-S data set}\label{data}
\subsection{Galaxy samples}\label{samples}
We work with the galaxy catalog from 
the Great Observatories Origins Deep Survey South (GOODS-S)\footnote{http://www.stsci.edu/science/goods/} field by the {\it Spitzer} Legacy Team
\citep{giavalisco04}. We used the Version 1.0
catalogs\footnote{http://archive.stsci.edu/prepds/goods/} and reduced mosaics in
the $F435W$ ($B_{435}$), $F606W$ ($V_{606}$), $F775W$ ($i_{775}$), and $F850LP$
($z_{850}$) {\it Hubble Space Telescope}/Advanced Camera for Surveys ({\it HST}/ACS) bands. These catalogs were cross-correlated using a
$1.5^{\prime\prime}$ search radius with the GOODS-S Infrared Array Camera (IRAC) selected sample in the
Rainbow cosmological database\footnote{http://guaix.fis.ucm.es/$\sim$pgperez/Proyectos/ \\ucmcsdatabase.en.html} published in \citet[see also \citealt{pgon05} and G.
Barro et al., in preparation]{pgon08}, which provided us spectral energy
distributions (SEDs) in the UV-to-MIR range, well-calibrated and with reliable
photometric redshifts, stellar masses, star formation rates, and rest-frame
absolute magnitudes.

We refer the reader to the mentioned papers for a more detailed description of
the data included in the SEDs and the analysis procedure. Here, we summarize
briefly the main characteristics of the data set. The Rainbow database contains consistent
aperture photometry in several UV, optical, NIR, and MIR bands with the method
described in \citet{pgon08}. UV-to-MIR SEDs were built for $4927$ IRAC sources in the GOODS-S region down to a 75\% completeness magnitude $[3.6]$$=$23.5~mag
(AB). These SEDs were fitted to stellar population and dust emission models to
obtain estimates of the photometric redshift ($z_{\rm phot}$), the stellar
mass ($M_{\star}$), and the rest-frame B-band absolute magnitude ($M_B$).

Rest-frame absolute $B$-band magnitudes were
estimated for each source by convolving the templates fitting the SED with the
transmission curve of a typical Bessel-$B$ filter, taking into account the
redshift of each source. This procedure provided us accurate interpolated
$B$-band magnitudes including a robustly estimated $K$-correction. Stellar
masses were estimated using the exponential star formation PEGASE01 models with
a \citet{salpeter55} initial mass function (IMF) and various ages, metallicities and dust contents
included. The typical uncertainties in the stellar masses are a factor of $\sim$2, 
typical of most stellar population studies \cite[see, e.g.][]{papovich06,fontana06}.

The median accuracy of the photometric redshifts at $z < 1.5$ is $|z_{\rm spec}
- z_{\rm phot}|/(1+z_{\rm spec}) = 0.4$, with a fraction $<$5\% of catastrophic
outliers \citep[][Figure~B2]{pgon08}. In the present paper we use $\sigma_{z_{\rm phot}} =
\sigma_{\delta_z} (1+z_{\rm phot})$ as $z_{\rm phot}$ uncertainty, where
$\sigma_{\delta_z}$ is the standard deviation in the distribution of the
variable $\delta_z \equiv (z_{\rm phot} - z_{\rm spec}) / ({1 + z_{\rm phot}})$,
that is well described by a Gaussian with mean $\mu_{\delta_z} \sim 0$ and $\sigma_{\delta_z}$ (see \citealt{clsj09ffgs}, for details).
We take
$\sigma_{\delta_z} = 0.043$ for $z \leq 0.9$ sources and $\sigma_{\delta_z} =
0.05$ for $z > 0.9$ sources. 

From the Rainbow catalog described above we defined two samples in the range $0.1~\leq~z~<~1.3$. One sample is selected in
luminosity, $M_B \leq -20$ ($\sim M_{B}^{*}$ at $z \sim 0$; \citealt{faber07}), which comprises 1122 sources. The value of $M_{B}^{*}$ is 1 mag brighter at $z \sim 1$ than locally \citep{gabasch04,ilbert05,faber07}, evolution that reflects the descent in the star formation rate density of the universe since $z \sim 1$ \citep[e.g.][]{hopkins06sfr}. Hence, if we assume a constant $M_B$ cut, we select different areas of the luminosity function at each redshift, biasing our results. However, L09 show that asymmetry as reliable morphological indicator in GOODS-S is valid only for $M_B \leq -20$ galaxies, so we decided to use this constant cut, instead of an evolving one, to ensure good statistics. This selection condition our results, so is important take it into account when we interpret the evolution of the comoving number density of galaxies with redshift (Section~\ref{netlt}). The second sample is selected in mass, 
$M_{\star} \geq 10^{10}\ M_{\odot}$ ($\sim 0.1M_{\star}^{*}$ at $z \sim 0$; \citealt{pgon08}), which comprises 987 galaxies. In this case $M_{\star}^{*}$ evolves little, if any, since $z \sim 1$ \citep{pozzetti07,pgon08,mascherini09}, so we decide to use a constant selection in mass that ensures 75\% completeness for passively evolving galaxies in the range under study (see L09, for details).

\subsection{Morphological Indices}\label{indices}
We use concentration ($C$; \citealt{abraham94}) and asymmetry ($A$;
\citealt{abraham96}) indices to perform our structural study. Concentration is
defined as
\begin{equation}
C = 5\times\log\bigg(\frac{r_{80}}{r_{20}}\bigg),
\end{equation}
where $r_{20}$ and $r_{80}$ are the circular radii which contain 20\% and 80\%
of the total galaxy flux, respectively. Concentration correlates with several
properties of galaxies, as bulge-to-total ratio \citep{strateva01,conselice03},
absolute $B$-band magnitude \citep{conselice03}, or stellar mass
\citep{conselice06me}. For details about concentration measurements see
\citet{bershady00}.

The asymmetry index is defined as
\begin{equation}
A = \frac{\sum |I_0 - I_{180}|}{\sum |I_0|} - \frac{\sum |B_0 - B_{180}|}{\sum
|I_0|},\label{A}
\end{equation}
where $I_0$ and $B_0$ are the original galaxy and background images, $I_{180}$
and $B_{180}$ are the original galaxy and background images rotated $180^{\circ}$,
and the summation spans all the pixels of the images. The background image is a
sky source-free section of $50 \times 50$ pixels. This index gives us information over
the source distortions and, we can use it to identify recent merger systems which
are highly distorted
\citep[e.g.,][]{conselice03,depropris07,bridge07,clsj09ffgs}. All details about
the measurement of the asymmetry index in the {\it HST}/ACS images of GOODS-S sources are given in
L09, with similar techniques having been applied for the measurement of the $C$ index. To avoid
statistical morphological $K$-corrections and to deal with the loss of
information with redshift (i.e., spatial resolution degradation and cosmological
dimming), we determined $C$ and $A$ indices in $B$-band rest-frame galaxy
images, which were artificially redshifted to a unique, representative
redshift, $z_{\rm d} = 1$. This provides us
an homogeneous data set to perform structural studies in GOODS-S.

\begin{deluxetable*}{lcccc} 
\tabletypesize{}
\tablecolumns{5} 
\tablewidth{0pc} 
\tablecaption{Early-type fraction in GOODS-S at $0.6 \leq z <
0.85$\label{lsstab}}
\tablehead{
Sample Selection & $n_{\rm LSS}$ & w/ LSS\ \tablenotemark{a}& w/o LSS\
\tablenotemark{b} & LSS ($z = 0.735$)\tablenotemark{c}}
\startdata
$M_B \leq -20$  			& 72 & $0.397^{+0.029}_{-0.027}$ &
$0.332^{+0.033}_{-0.030}$ & $0.634^{+0.045}_{-0.051}$ \\
$M_{\star} \geq  10^{10}\ M_{\odot}$ 	& 94 & $0.509^{+0.029}_{-0.027}$ &
$0.439^{+0.035}_{-0.033}$ & $0.697^{+0.041}_{-0.047}$
\enddata 
\tablenotetext{a}{Early-type fraction in the sample {\it with LSS sources}.}
\tablenotetext{b}{Early-type fraction in the sample {\it without LSS sources}.}
\tablenotetext{c}{Early-type fraction in the LSS.}
\end{deluxetable*}

\subsection{Major merger rates}\label{merger}
We use the major merger rates from L09 thought this paper. In that work, we selected as major merger remnants those galaxies with high values of the asymmetry index ($A > 0.3$). We performed that study for $M_{\star} \geq 10^{10}\ M_{\odot}$ galaxies up to $z \sim 1$ in the same mass-selected sample that we use in the present paper (Section~\ref{samples}), and we took into account the effect of observational errors in $z$ and $A$, which overestimate the merger fraction due to the spillover of normal sources to the high-asymmetry regime, by maximum likelihood (ML) techniques developed in \citet{clsj08ml}. We obtain lower merger fractions ($f_{\rm m}^{\rm mph} \lesssim 5$\%) than other determinations also based on morphology \citep[e.g.,][]{lotz08ff,conselice09cos}, but are in good agreement with those from \citet{clsj09ffgs} in Groth Strip applying the same methodology as that of ours, and those from \citet{jogee09} in Galaxy Evolution from Morphology and SEDs\footnote{http://www.mpia.de/GEMS/gems.htm} (GEMS) by eye-ball inspection of the sources. To obtain the merger rate ($\Re_{\rm m}$) from the merger fraction, we assume \citet{pgon08} mass functions and a typical timescale of $T_A \sim 0.5$ Gyr \citep{conselice06ff,conselice09t,lotz08t,lotz09t}.

The methodology in L09 is only sensitive to gas-rich major mergers (i.e., at least one of the merging galaxies is a LT, gas-rich galaxy): asymmetry as merger indicator is calibrated in local universe with ultraluminous infrared galaxies \citep{conselice03}, while $T_A$ is determined by $N$-body simulations of gas-rich major mergers \citep{conselice06ff,lotz08t}. In a forthcoming paper, we show that nearly all the high-asymmetry galaxies in our sample are also star-forming galaxies, supporting the notion that we are only sensitive to gas-rich mergers (C. Lopez-Sanjuan et al, in preparation).

\section{The Concentration-Asymmetry Bimodality} \label{cabimo}
Concentration and asymmetry indices are useful to segregate galaxies by their structure, i.e., bulge-dominated galaxies (E/S0/Sa, ET galaxies in the
following), and disk-dominated galaxies and irregulars (Sb-Irr, LT
galaxies in the following;
\citealt{bershady00,lauger05,menanteau06,yagi06,huertas08m,neichel08}). To
segregate galaxies by their morphology we need an index as the cumpliness (S;
\citealt{conselice03}), or the bumpiness (B; \citealt{blakeslee06}), which give
us information about the distribution of the star formation in the galaxy
\citep{vanderwel08}. 
On the other hand, high values of the asymmetry index select
gas-rich major merger remnants (Section~\ref{merger}). 

\begin{figure}[t]
\plotone{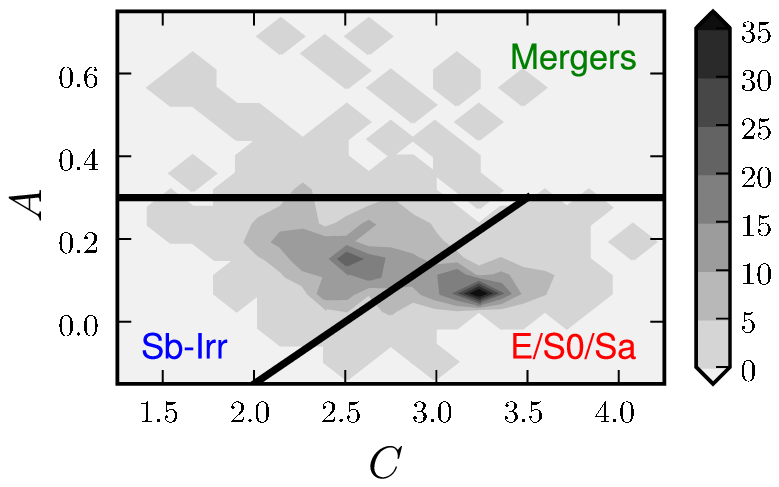}
\caption{Distribution in the concentration-asymmetry plane of the 757 galaxies
in the catalog with $M_B \leq -20$ and $0.35 \leq z < 1.1$. The black lines
mark the different structural areas: ET galaxies (E/S0/Sa), LT
galaxies (Sb-Irr), and gas-rich major mergers. The contours show number of
galaxies (see the right legend).}
\label{cabimofig}
\end{figure}

In Figure~\ref{cabimofig} we show the distribution of the 757 galaxies in the
catalog with $M_B \leq -20$ and $0.35 \leq z < 1.1$, which present two different
populations, in agreement with \citet{zamojski07} and \citet{conselice09cos}.
One population is associated with ET galaxies, with its maximum at
$(C,A) = (3.25,0.09)$, and the other with LT galaxies, with its
maximum at $(C,A) = (2.52,0.17)$, as shown by \citet{ilbert06}. We determined
the positions of the two peaks in $(C,A)$ space as follows: first we fitted the histogram in
concentration space with two Gaussians, which gave us the concentration
values of the two peaks as well as the minimum between them ($C = 2.9$). Then, we fitted the asymmetry
histogram of $C < 2.9$ and $C > 2.9$ galaxies with a simple Gaussian to obtain
the asymmetry maxima. In Figure~\ref{cabimofig} we also show the three areas
that
we use to segregate galaxies: gas-rich major mergers ($A
> 0.3$, see L09), ET galaxies ($A < 0.3 \cap A < 0.3C -
0.75$), and LT galaxies ($A < 0.3 \cap A > 0.3C - 0.75$). The limit $A =
0.3C - 0.75$ was chosen to optimize the ET/LT separation (see the following
section for details).

\subsection{Early- and Late-Type Fraction Determination}\label{metodo}
The first goal of this paper is to study the fraction of ET ($f_{\rm ET}$) and LT galaxies ($f_{\rm LT}$) as a function of redshift. To obtain reliable ET and LT fractions we used ML techniques.  
\citet{clsj08ml} developed a two-dimensional ML method to determine merger fractions from the asymmetry in galaxy images; the method is far superior to classical source counting, as, with source counting, measurement errors cause source spillover to neighboring bins.  
The ML method has been successfully used to measure the
gas-rich merger fraction in Groth strip \citep{clsj09ffgs} and in GOODS-S
(L09), and we apply it here for the determination of $f_{\rm ET}$ and $f_{\rm LT}$. We refer the reader to \cite{clsj08ml} for all the details about the ML
method and its assumptions. The main difficulty in the application of the ML method in the present study is that it works in two dimensions, whereas our problem has three dimensions: $C$, $A$, and the redshift $z$. 

We could work in ($C,z$) space, and use as selection limit the
condition $C = 2.9$, which marks the separation between ET and LT
populations in concentration space (see the previous section). However, this selects
galaxies with $A > 0.3$, i.e., gas-rich major mergers, as ET and LT
galaxies. Indeed, the separation between ETs and LTs is oblique in Figure~\ref{cabimofig} and is well described by $A = 0.3C - 0.75$ (drawn in the figure).  
If we define a second variable $CA \equiv A - 0.3C +
0.75$, LT galaxies {\it and} major mergers have $CA > 0$, while
ET galaxies have $CA < 0$. That is, the fraction of
ET galaxies is $f_{\rm ET} = f_{CA < 0}$, while the fraction of
LT galaxies is $f_{\rm LT} = f_{CA > 0} - f_{\rm m}^{\rm mph}$, where
$f_{\rm m}^{\rm mph}$ is the morphological merger fraction reported by L09. The
error in the new variable $CA$ is $\sigma^2_{CA} = \sigma^{2}_{A} +
0.09\sigma^{2}_{C}$.

How are the results affected by the selection limit? We determined $f_{\rm ET}$
with three different limits: $C = 2.9$, $A = 0.3C - 0.75$, and $A = \frac{5}{6}C
- 2.30$. We find that all the values of $f_{\rm ET}$ are consistent within
$1\sigma$ with our preferred condition, $A = 0.3C - 0.75$, in the three redshift
ranges under study, namely, $z_1 = [0.35,0.6), z_2 = [0.6,0.85)$, and $z_3 =
[0.85,1.1)$. We select these particular ranges to resemble those in L09. Because the particular election of the selection limit does not
bias our results, we use the limit $A = 0.3C - 0.75$ in the following. 

\subsection{Large-Scale Structure Effect}\label{lss}
\begin{figure}[t!]
\plotone{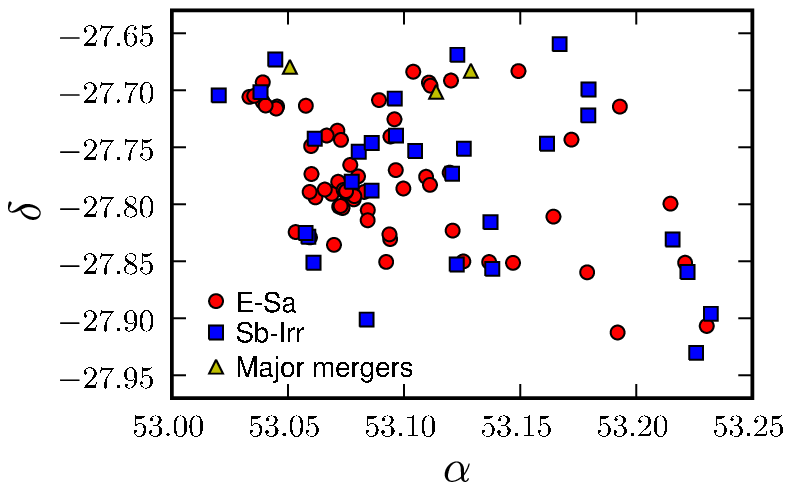}
\caption{Position in the sky plane of the $M_{\star} \geq 10^{10}\ M_{\odot}$
galaxies in the more prominent LSS in GOODS-S field ($z = 0.735$). Red dots are ET galaxies, blue squares are LT galaxies, and
green triangles are gas-rich major mergers.}
\label{lssfig}
\end{figure}

It is well known that the more prominent large-scale structure (LSS) in the
GOODS-S field is located at redshift $z = 0.735$ \citep{ravikumar07}. The next two more important ones are located at $z = 0.66$ 
and $z = 1.1$. The former is an overdensity in redshift space, but not in the sky plane, 
while the latter is a cluster, but comprises an order of magnitude fewer sources than the
 $z = 0.735$ structure (145 versus 12, \citealt{Adami05}). Because of this, we concentrate on the LSS at $z = 0.735$ and ignore other structures in GOODS-S. In order
to check the effect of this LSS in our derived structural fractions, we
recalculated the ET fraction in the range $z_2 = [0.6,0.85)$ by
excluding the sources within $\delta v \leq 1500\ {\rm km\ s}^{-1}\ (\delta z
\sim 0.01)$ of $z = 0.735$ (\citealt{rawat08}; L09). In Table~\ref{lsstab}, we
summarize the number of sources in the LSS for each sample ($n_{\rm LSS}$), the ET fractions in the LSS and the ET fractions in the field, both in the samples with and without LSS sources. The value of $f_{\rm ET}$ in the samples without LSS sources is $\sim$0.07 lower than the determination in the whole samples. More importantly,
both values are incompatible at $1\sigma$. This fact explains the high ET
fraction found by \citet{lauger05me} at that redshift in their GOODS-S field
study. Because of this, in the following we use the structural fractions
obtained from the samples {\it without LSS sources}. On the other hand, $f_{\rm
ET}^{\rm LSS} \sim 1.8 f_{\rm ET}^{\rm field}$, that is, the ET fraction
is higher in dense environments. In Figure~\ref{lssfig}, we show the position in
the sky plane of ET (red dots), LT (blue squares), and major
mergers (green triangles) for $M_{\star} \geq 10^{10}\ M_{\odot}$ galaxies in the
LSS: ET galaxies are concentrated in more populated regions, while
LT galaxies are located in the outskirts of the structure, as expected by the
morphology-density relation \citep{dressler80}. In addition, the three gas-rich
major mergers in the LSS are also located in the outskirts, a natural
consequence of the external location of their progenitors, i.e. LT
galaxies (see also \citealt{heiderman09}).

\begin{figure}[t!]
\plotone{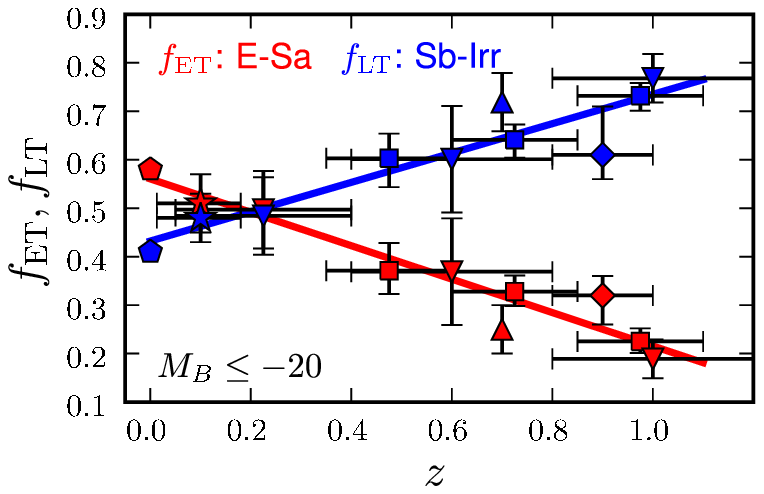}
\plotone{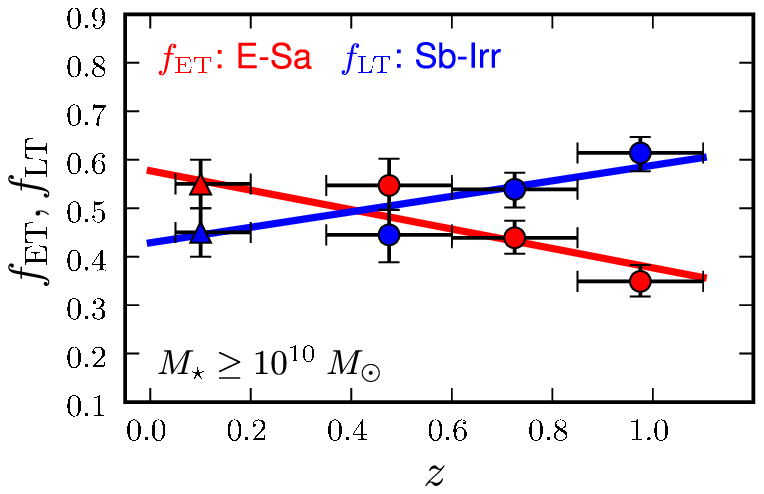}
\caption{ET (red) and LT galaxy fraction (blue) vs. redshift. {\bf{\it Top}}: structural fractions for $M_B \leq -20$ galaxies: this work (squares),
\citet[][diamonds]{lotz08ff}, \citet[][inverted triangles]{ilbert06},
\citet[][triangles]{scarlata07me}, \citet[][stars]{driver06}, and \citet[][pentagons]{conselice06me}. {\bf{\it
Bottom}}: structural fractions for $M_{\star} \geq 10^{10}\ M_{\odot}$ galaxies:
this work (circles), and \citet[][triangles]{mandelbaum06}. In both panels, the
redshift errors show the redshift range covered for each data, while red (blue) line is the linear least-squares fit to the $f_{\rm ET}$ ($f_{\rm LT}$) data.}.
\label{ffig}
\end{figure}

\begin{deluxetable}{lcc} 
\tabletypesize{}
\tablecolumns{3} 
\tablewidth{0pc} 
\tablecaption{Structural fractions of $M_B \leq -20$ galaxies\label{fbtab}}
\tablehead{ 
$z$ & $f_{\rm ET}$ & $f_{\rm LT}$}
\startdata
$0.475$ & $0.371^{+0.057}_{-0.048}$ & $0.603^{+0.051}_{-0.059}$ \\
$0.725$ & $0.332^{+0.033}_{-0.030}$ & $0.637^{+0.032}_{-0.037}$ \\
$0.975$ & $0.225^{+0.027}_{-0.024}$ & $0.732^{+0.026}_{-0.031}$
\enddata 
\end{deluxetable}

\section{The structural evolution of GOODS-S galaxies}\label{results}
We summarize in Table~\ref{fbtab} the structural fractions for $M_B \leq -20$
galaxies in three redshift ranges: $z_1 = [0.35,0.6), z_2 = [0.6,0.85)$, and
$z_3 = [0.85,1.1)$. The fraction of ET galaxies increases with cosmic
time, while LT galaxies fraction decreases. Note that errors are formal
and do not take into account cosmic variance, denoted $\sigma_{v}$;
following \citet{somerville04}, we expect $\sigma_{v}\sim20$\%. In the top panel of
Figure~\ref{ffig}, we show our results (squares) with those from the literature:
diamonds are the ET and LT fractions provide by \citet{lotz08ff} for
$M_B \leq -18.8 -1.3z$ galaxies in All-Wavelength Extended Groth Strip International Survey (AEGIS\footnote{http://aegis.ucolick.org/}; \citealt{davis07}),
which mimic our selection criteria at $z \sim 0.9$. Pentagons are from
\citet{conselice06me} for $M_B \leq -20$ galaxies in the
RC3\footnote{http://heasarc.nasa.gov/W3Browse/all/rc3.html} (Third Reference
Catalogue of Bright Galaxies; \citealt{deva91}): we take ET = E/S0 and
early-disks, while LT = late-disks and irregulars (his Table~1). Triangles and
inverted triangles are obtained integrating the morphological luminosity
functions (MLF) provided by \citet{scarlata07me} at $z = 0.7$, and
\citet{ilbert06} at $0.05 < z < 1.2$, respectively. In both cases, we integrated
the MLF for galaxies brighter than $M_B = -20$. \citet{ilbert06} perform their
study in the GOODS-S field and segregate galaxies in ET/LT by their
positions in the $C--A$ plane, a similar methodology to ours. \citet{scarlata07me}
use Zurich Estimator of Structural Types (ZEST) in Cosmological
Evolution Survey (COSMOS\footnote{http://cosmos.astro.caltech.edu/index.html}; \citealt{scoville07})
to classify $\sim$10,000 galaxies in several morphological types: E/S0 (T1), disks (T2, divided in four subtypes by
their S\'ersic index) and irregulars (T3). We take ET = T1 + T2.0, while LT =
T2.1 + T2.2 + T2.3 + T3 $ - f_{\rm m}^{\rm mph}(0.7)$, where $f_{\rm m}^{\rm
mph}(0.7)$ is the morphological merger fraction at $z = 0.7$ from L09. This
selects disks with the S\'ersic index $n > 2.5$ as bulge-dominated galaxies \citep{mandelbaum06,trujillo07,dahlen07}. Finally, stars are obtained integrating the $n \geq 2$ (ETs) and $n < 2$ (LTs) luminosity functions provided by \citet{driver06} from Millennium Galaxy Catalog (MGC\footnote{http://www.eso.org/$\sim$jliske/mgc/}; \citealt{liske03}). They found that their $z \sim 0.1$ galaxies present a bimodality in $n$, with a minimum at $n \sim 2$. In the three previous cases we determine the uncertainty in structural fractions by varying 1$\sigma$, the value of $M^{*}_B$ in the integration. The top panel of Figure~\ref{ffig} shows that, when we use a
similar luminosity and structural selection, all $f_{\rm ET}$ and $f_{\rm LT}$
are in good agreement. The linear least-squares fit to the data yields
\begin{eqnarray}
f_{\rm ET}(z) = (0.56\pm0.02) - (0.35\pm0.02)z,\\
f_{\rm LT}(z) = (0.43\pm0.02) + (0.30\pm0.03)z.
\end{eqnarray}
With these fits, we infer that ET galaxies are the dominant
$M_B \leq -20$ population by number since $z \sim 0.2$.

\begin{deluxetable}{lcc} 
\tabletypesize{}
\tablecolumns{3} 
\tablewidth{0pc} 
\tablecaption{Structural fractions of $M_{\star} \geq 10^{10}\ M_{\odot}$
galaxies\label{fstab}}
\tablehead{ 
$z$ & $f_{\rm ET}$ & $f_{\rm LT}$}
\startdata
$0.475$ & $0.547^{+0.055}_{-0.050}$ & $0.445^{+0.051}_{-0.056}$ \\
$0.725$ & $0.439^{+0.035}_{-0.033}$ & $0.539^{+0.034}_{-0.037}$ \\
$0.975$ & $0.349^{+0.034}_{-0.031}$ & $0.614^{+0.033}_{-0.038}$
\enddata 
\end{deluxetable}

In Table~\ref{fstab} we summarize the structural fractions for $M_{\star} \geq
10^{10}\ M_{\odot}$ galaxies. The fraction of ET galaxies
increases with cosmic time, as in the previous case, but $f_{\rm ET}$ in the
mass-selected sample is higher than in the luminosity-selected sample in all the
redshift intervals under study. In the bottom panel of Figure~\ref{ffig} we show our
results (circles) with those from \citet[][triangles]{mandelbaum06}. They
provide $f_{\rm ET}$ ($n > 2.5$) and $f_{\rm LT}$ ($n < 2.5$) for $\sim$33000
$M_{\star} \geq 10^{10}\ M_{\odot}$ galaxies from Sloan Digital Sky Survey
(SDSS\footnote{http://www.sdss.org/}; \citealt{adelman06}). The linear least-squares fit to the data yields
\begin{eqnarray}
f_{\rm ET}(z) = (0.60\pm0.04) - (0.24\pm0.06)z,\label{fets}\\
f_{\rm LT}(z) = (0.40\pm0.04) + (0.19\pm0.06)z.\label{flts}
\end{eqnarray}
With these fits, we infer that ET galaxies are the dominant $M \geq 10^{10}\ M_\odot$ 
population by the number since $z \sim 0.5$, a higher redshift than in the
luminosity-selected sample. Although in the range of study a linear function is a good approximation to the observed evolution, we expect that a power-law function may provide a better parameterization when higher redshift data are available \citep[e.g.][]{fontana09}.   

\subsection{Comoving Number Density Evolution}\label{netlt}
To better understand the structural evolution of galaxies since $z = 1$, in this
section we study the comoving number density of ET ($\rho_{\rm ET}$) and
LT galaxies ($\rho_{\rm LT}$) as a function of redshift:
\begin{eqnarray}
\rho_{\rm ET}(z,M) = f_{\rm ET}(z,M)\rho_{\rm tot}(z,M),\label{neteq}\\
\rho_{\rm LT}(z,M) = f_{\rm LT}(z,M)\rho_{\rm tot}(z,M),\label{nlteq}
\end{eqnarray}
where $M = M_B$ $[M_{\star}]$ is the selection criteria, and $\rho_{\rm tot}(z,M)$ is the
comoving number density at redshift $z$ of galaxies more luminous [massive] than
$M_B$ $[M_{\star}]$. To obtain $\rho_{\rm tot}(z,M_B)$, we assume the \citet{faber07}
luminosity function parameters, while to obtain $\rho_{\rm tot}(z,M_{\star})$ we assume
the \citet{pgon08} mass function parameters.

\begin{figure}[t!]
\plotone{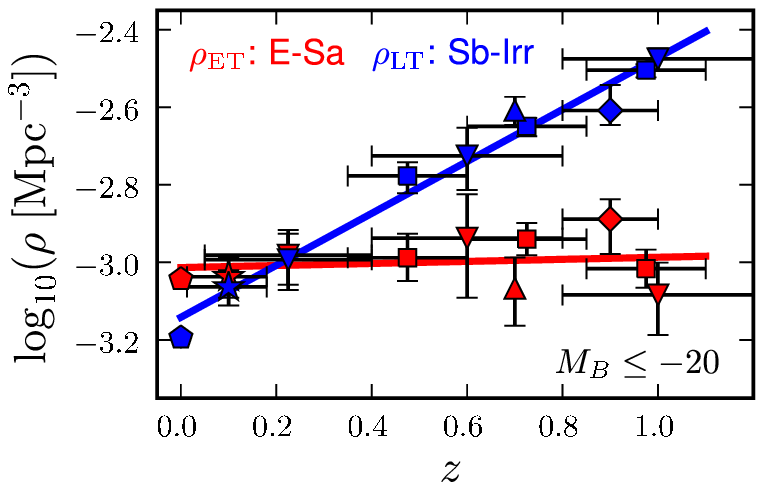}
\plotone{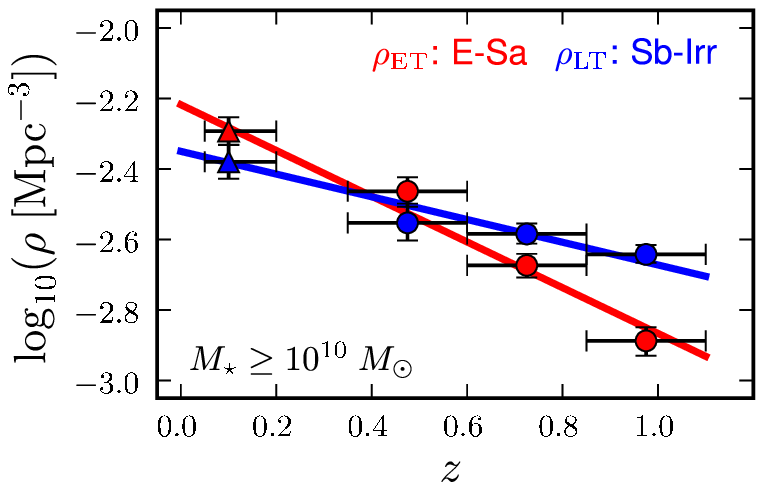}
\caption{ET (red) and LT comoving number density (blue) vs. redshift. {\bf{\it Top}}: structural number density for $M_B \leq -20$ galaxies: this work
(squares), \citet[][diamonds]{lotz08ff}, \citet[][inverted triangles]{ilbert06},
\citet[][triangles]{scarlata07me}, \citet[][stars]{driver06}, and \citet[][pentagons]{conselice06me}. {\bf{\it Bottom}}: structural fractions for $M_{\star} \geq 10^{10}\ M_{\odot}$ galaxies:
this work (circles), and \citet[][triangles]{mandelbaum06}. In both panels the
redshift error bars show the redshift range covered for each data, while red (blue) line is the linear least-squares fit to the $\rho_{\rm ET}$ ($\rho_{\rm LT}$) data.}
\label{nfig}
\end{figure}

\begin{deluxetable}{lccc} 
\tabletypesize{}
\tablecolumns{4} 
\tablewidth{0pc} 
\tablecaption{Comoving number densities of $M_B \leq -20$ galaxies in GOODS-S\label{nbtab}}
\tablehead{
$z$ & $\rho_{\rm ET}$ & $\rho_{\rm LT}$ & $\rho_{\rm tot}(M_B)$\\
    & $(10^{-3}\ {\rm Mpc^{-3}})$ & $(10^{-3}\ {\rm Mpc^{-3}})$ & $(10^{-3}\ {\rm Mpc^{-3}})$}
\startdata
$0.475$ & $1.02 \pm 0.15$ & $1.73 \pm 0.15$ & 2.77\\
$0.725$ & $1.16 \pm 0.10$ & $2.26 \pm 0.12$ & 3.50\\
$0.975$ & $0.96 \pm 0.11$ & $3.16 \pm 0.12$ & 4.28
\enddata 
\end{deluxetable}

We summarize in Table~\ref{nbtab} the comoving number density for $M_B \leq -20$
galaxies. The number density of LT galaxies decreases with cosmic time,
while ET number density is roughly constant. In the top panel of
Figure~\ref{nfig}, we show the number density data of this work and those from
literature, obtaining by applying Equation~(\ref{neteq}) and (\ref{nlteq}) to the
fractions in Figure~\ref{ffig}. The linear least-squares fit to all data yields
\begin{eqnarray}
\log_{10}(\rho_{\rm ET}) = (-3.01 \pm 0.03) + (0.03 \pm 0.04)z,\label{netb}\\
\log_{10}(\rho_{\rm LT}) = (-3.14 \pm 0.02) + (0.67 \pm 0.02)z.\label{nltb}
\end{eqnarray}
The number density of LT galaxies at $z = 0$ is $\sim 20$\% the value at 
$z = 1$, while $\rho_{\rm ET} \sim$ constant.\\

\begin{deluxetable}{lccc} 
\tabletypesize{}
\tablecolumns{4} 
\tablewidth{0pc} 
\tablecaption{Comoving number densities of $M_{\star} \geq 10^{10}\ M_{\odot}$ galaxies in
GOODS-S\label{nstab}}
\tablehead{
$z$ & $\rho_{\rm ET}$ & $\rho_{\rm LT}$ & $\rho_{\rm tot}(M_\star)$\\
    & $(10^{-3}\ {\rm Mpc^{-3}})$ & $(10^{-3}\ {\rm Mpc^{-3}})$ & $(10^{-3}\ {\rm Mpc^{-3}})$}
\startdata
$0.475$ & $3.41 \pm 0.33$ & $2.82 \pm 0.36$ & 6.29\\
$0.725$ & $2.12 \pm 0.16$ & $2.60 \pm 0.17$ & 4.83\\
$0.975$ & $1.28 \pm 0.12$ & $2.29 \pm 0.13$ & 3.71
\enddata 
\end{deluxetable}

The {\it total} number density of $M_B \leq -20$ galaxies decreases by a factor of 3 since $z = 1$, a reflection of the global decline in the star formation rate density
of the universe as $\sim(1+z)^{4}$ \citep[e.g.][]{lilly96,hopkins06sfr,pgon05,tresse07,villar08}. Because both the number densities of LT galaxies and that of gas-rich major mergers decrease with cosmic time (e.g. L09), we ask which of these two populations drives the decrease in the star formation rate density of the universe.  

We define the ratio
between the star formation rate in a given population (SFR$_{\rm pop}$) and the
star formation rate in major merger systems (SFR$_{\rm m}$) as:
\begin{equation}
R_{\rm pop} \equiv \frac{{\rm SFR}_{\rm pop}}{{\rm SFR}_{\rm m}} = \frac{\mu_{\rm pop}f_{\rm
pop}}{\epsilon_{{\rm SF}}f^{\rm mph}_{\rm m}}\frac{T_{{\rm
m},A}}{T_{\epsilon}},\label{rpop}
\end{equation}
where $f^{\rm mph}_{\rm m}$ and $f_{\rm pop}$ are the morphological major merger
fraction (i.e., gas-rich, Section~\ref{merger}), and the fraction of galaxies of a given population, respectively,
$\mu_{\rm pop}$ is the fraction of galaxies in the population that are forming stars
actively, $\epsilon_{{\rm SF}}$ is the star formation in an interacting system respect
to the median in isolated galaxies, $T_{\epsilon}$ is the time over which the
star formation is enhanced in an interaction, and $T_{{\rm m},A}$ is the
timescale over which one major merger is selected as high-asymmetric source by L09 methodology. In Equation~(\ref{rpop}), we have assumed that all the
gas-rich major mergers have enhanced star formation, and we do not take into account dry mergers, that have negligible star formation. With this definition, the
total star formation rate (SFR$_{\rm tot}$) in a given redshift range is
\begin{equation}
{\rm SFR}_{\rm tot} = (R_{\rm LT} + R_{\rm ET} + 1){\rm SFR}_{\rm m},
\end{equation}
while the fraction of the total star formation located in major merger systems
is
\begin{equation}
f_{{\rm SF},{\rm m}} = \frac{1}{R_{\rm LT} + R_{\rm ET} + 1}.
\end{equation}
Because most of the red galaxies are ETs (\citealt{lotz08ff}), we assume
that $\mu_{\rm ET} = 0$ (i.e., all ET galaxies are passive), and $R_{\rm
ET} = 0$. On the other hand, most of the blue galaxies are LTs
(\citealt{lotz08ff}), so we take $\mu_{\rm LT} = 1$. With these assumptions the
star formation rate is more important in LT galaxies than
in major merger systems if $R_{\rm LT} > 1$ and vice versa if $R_{\rm LT} < 1$. To
compare the results with previous works, we restrict our study to the range $0.4
\leq z < 0.8$, in which $f^{\rm mph}_{\rm m} = 0.026^{+0.014}_{-0.009}$ and
$f_{\rm LT} = 0.610^{+0.028}_{-0.032}$. We assume that $T_{{\rm m},A} = 0.475
\pm 0.125$ Gyr (L09), $\epsilon_{\rm SF} = 1.50\pm0.25$ (\citealt{robaina09}, see
also \citealt{lin07,li08,knapen09}), and $T_{\epsilon} = 2.0 \pm 0.25$ Gyr
\citep{dimatteo07,cox08}. With these values we obtain $R_{\rm LT} =
3.7^{+1.9}_{-2.5}$, that is, {\it the bulk of the star formation in the range
$0.4 \leq z < 0.8$ is located in LT galaxies}, and major mergers account
for $\sim$20\% of the total star formation rate, in quantitative agreement with
\citealt{bell05}, \citealt{wolf05sf}, \citealt{jogee09}, or \citet{sobral09}.

Following \citet{robaina09}, we can also obtain the fraction of SFR$_{\rm tot}$
that is triggered directly by mergers:
\begin{equation}
f^{\rm tri}_{{\rm SF},{\rm m}} = \frac{\epsilon_{\rm SF}-1}{\epsilon_{\rm SF}}f_{{\rm SF},{\rm m}} =
\frac{\epsilon_{\rm SF}-1}{\epsilon_{\rm SF}(R_{\rm LT} + R_{\rm ET} + 1)}.
\end{equation}
If we assume that $R_{\rm ET} = 0$, as previously, then $f^{\rm tri}_{{\rm SF},{\rm
m}} = 7^{+8}_{-2}$\%, a low value in good agreement with \citet{robaina09}:
they infer that $f^{\rm tri}_{{\rm SF},{\rm m}} = 8\pm3$\% in
GEMS\footnote{http://www.mpia-hd.mpg.de/GEMS/gems.htm} \citep{rix04}, at the same redshift range, and by
studying the correlation function of $M_{\star} \geq 10^{10}\ M_{\odot}$ blue
galaxies. If we repeat this study in the range $0.85 \leq z < 1.1$, we obtain
$R_{\rm LT} = 2.4$ (i.e., $\sim 30$\% of the total star formation is located in major mergers at $z \sim 1$), and $f^{\rm tri}_{{\rm SF},{\rm m}} \sim 9$\%. Our values are also in good agreement with \citet{hopkins09sfr}; they inferred the merger-induced burst history up to $z \sim 2$ from the luminosity profiles of local E/S0 galaxies, finding that $\sim 5$\%--10\% of total star formation is triggered by mergers, independently of redshift. Finally, our results are the same if we use the mass-selected sample.\\

In Table~\ref{nstab}, we summarize the number densities for $M_{\star} \geq
10^{10}\ M_{\odot}$ galaxies. In this case the behavior is the opposite than in
the luminosity-selected sample: the number densities of ET and LT
galaxies {\it increase} with cosmic time. This increase is more important for
ET galaxies. In the bottom panel of Figure~\ref{nfig}, we show the number
density data of this work (circles) and those from
\citet[][triangles]{mandelbaum06}. The linear least-squares fits to the data
are
\begin{eqnarray}
\log_{10}(\rho_{\rm ET}) = (-2.19 \pm 0.04) - (0.68 \pm 0.06)z,\\
\log_{10}(\rho_{\rm LT}) = (-2.37 \pm 0.03) - (0.29 \pm 0.05)z.
\end{eqnarray}
The ET population increases its number density by a factor of 5 between
$z=1$ and $z=0$, while LT galaxies increase by a factor of 2 in the same
range. Because $\sim$40\% of $M_{\star} \geq 10^{10}\ M_{\odot}$ galaxies in the
local universe are in place at $z = 1$ \citep{pgon08}, the increase in the
number density of these galaxies since $z = 1$ is primarily due to the increase
in $\rho_{\rm ET}$. As in the previous section, we
expect that a power-law function may provide a better parameterization than the linear
when higher redshift data are available \citep[e.g.,][]{taylor09}. If we fit a power-law function to the
$\rho_{\rm ET}$ data, $\rho_{\rm ET}(z) \propto
(1+z)^{\alpha}$, we obtain $\alpha = -2.3\pm0.2$, while \citet{taylor09} find
$\alpha = -1.70\pm0.14$ for $M_{\star} \gtrsim 10^{11}\ M_{\odot}$ red sequence
galaxies. If we assume that most red sequence galaxies are ETs
\citep{lotz08ff}, this implies that red massive galaxies evolve less in number
density than less massive ones since $z \sim 1$, in agreement with
\citet{ferreras09}.

The results of this section suggest that {\it we need a structural
transformation in the range $0 < z < 1$ between LT galaxies, which
form stars actively, and ET galaxies, in which stellar mass is
located}, in agreement with \citet{bell07}, \citet{vergani08}, or \citet{ruhland09}. The question
is, can gas-rich major mergers drive this structural transformation?

\section{The role of gas-rich major mergers in the structural evolution of
intermediate-mass galaxies}\label{rolewet}
To explore the role of gas-rich major mergers in the structural evolution of
$M_{\star} \geq 10^{10}\ M_{\odot}$ galaxies we define the fraction of new ETs that appears between $z_2$ and $z_1$ due to gas-rich major mergers, $f_{\rm
ET,m}$, as:
\begin{equation}
f_{\rm ET,m}(z_1,z_2) \equiv \frac{\rho_{\rm ET,m}(z_1,z_2)}{\rho_{\rm ET}^{\rm
new}(z_1,z_2)} = \frac{\rho_{\rm rem}(z_1,z_2)}{\rho_{\rm ET}(z_1) - \rho_{\rm
ET}(z_2)},\label{fremetm}
\end{equation}
where $\rho_{\rm ET,m}(z_1,z_2)$ is the number density of new ETs due to
gas-rich mergers, and $\rho_{\rm ET}^{\rm new}(z_1,z_2)$ is the {\it total} number
density of new ETs. We assume that each gas-rich major merger remnant
is an ET galaxy \citep{naab06SS,rothberg06a,rothberg06b,hopkins08ss,hopkins09disk}, so $\rho_{\rm
ET,m}(z_1,z_2) = \rho_{\rm rem}(z_1,z_2)$, being
\begin{equation}
\rho_{\rm rem}(z_1,z_2) = \int_{z_1}^{z_2}\Re_{\rm m}(0)(1+z)^{n-1}\frac{{\rm
d}z}{H_0 E(z)},
\end{equation}
where $E(z) = \sqrt{\Omega_{\Lambda} + \Omega_{M}(1+z)^3}$ in a flat universe,
$\Re_{\rm m}(0) = (0.3\pm0.1) \times 10^{-4}\ {\rm Mpc^{-3}\ Gyr^{-1}}$, and $n
= 3.5 \pm 0.4$ (merger rate parameters are those from L09). Using the
results of
the previous section, we infer that $f_{\rm ET,m}(0,1) = 17^{+10}_{-7}$\%. The
errors take into account the uncertainties in the merger rate and number density
parameters. This value implies that {\it gas-rich major mergers cannot explain
the observed structural evolution}.

Interestingly, if we extrapolate the observed tendencies up to $z_2 = 1.5$, we
obtain that $f_{\rm ET,m}(1,1.5) \sim 100$\%. This is, gas-rich major mergers
could be the dominant process in ETs formation at $z \gtrsim 1$. Although extrapolating
our results beyond $z = 1$ is risky and more data in the range $1 < z < 2$ are
needed to explore the suggested picture, the very different behavior in the two
redshift intervals makes our results qualitatively reliable. 

How are our results modified by spheroid-spheroid (dry) mergers?
The L09 methodology is only sensitive to gas-rich mergers (Section~\ref{merger}), so the role of
dry mergers cannot be measured directly in this work. However, we can estimate the effect
of dry mergers in our results: if we assume that a dry merger between
two ET galaxies leads to another ET galaxy
\citep{gongar03,gongar06,naab06ee,boylan06,hopkins09ee}, then each dry merger remnant represents the disappearance of one ET galaxy. That is, the value of $\rho_{\rm
ET}^{\rm new}(z_1,z_2)$ in Equation~(\ref{fremetm}) could be higher due to dry mergers. In addition, some of our observed gas-rich mergers may be spiral--spheroid (mixed) mergers that conserve the number of ET galaxies (i.e., only spiral--spiral wet mergers create new spheroids). Because of this, the value of $\rho_{\rm rem}(z_1,z_2)$ in Equation~(\ref{fremetm}) could be lower due to mixed mergers.  In summary, our $f_{\rm ET,m}(0,1)$ value is at least an upper limit to the importance of gas-rich major mergers in the structural evolution of $M_{\star} \geq 10^{10}\ M_{\odot}$ galaxies.\\

\subsection{Comparison with Previous Studies}
In this section, we compare our result with those in literature. \citet{bundy09} perform a close pair study in GOODS-South and North, and compare the formation rate of new spheroids against their measured gas-rich (wet + mixed) merger rate. From their results, we infer $f_{\rm ET,m}(0.4,0.9) \sim 16^{+24}_{-10}$\% for $\log(M_{\star}/M_{\odot}) \sim 10.7$ galaxies\footnote{For our cosmology and assuming a \citet{salpeter55} IMF.} that compares well with our value in the same redshift range, $f_{\rm ET,m}(0.4,0.9) = 26^{+15}_{-10}$\%, for the main galaxy in our mass sample, which has $\log(M_{\star}/M_{\odot}) \sim 10.6$. Although we include Sa galaxies in our definition of ET galaxy, the conclusion from both works is similar: less than half of the new intermediate-mass ET that appeared at $0.4 < z < 0.9$ come from gas-rich major mergers. 

In their work, \citet{wild09} studied post-starburst (PSB) galaxies in COSMOS and their role in red sequence assembly. The spectra of PSB indicate that the formation of O- and early-B-type stars has suddenly ceased in the galaxy, while the simulations performed by \citet{johansson08} find that the PSB phase can only be reached by gas-rich major merger remnants. In L09, we show that the gas-rich merger rate and the PSB rate of $M_{\star} \gtrsim 10^{10}\ M_{\odot}$ galaxies at $0.5 < z < 1.0$ are similar, supporting PSB galaxies as descendants of our gas-rich mergers. In addition, most of the red sequence galaxies are ETs \citep[e.g.][]{lotz08ff}, so we can compare the fraction of red sequence mass from PSB galaxies at $0.5 < z < 1.0$, $38^{+4}_{-11}$\%, with $f_{\rm ET,m}(0.5,1) = 34^{+17}_{-13}$\%. Despite the different methodologies and assumptions, the agreement between both studies is remarkable.

Other works also study the impact of mergers in ETs evolution. \citet{bundy07} compare the evolution of the virial mass functions of halos hosting spheroidals with the merger rates predicted by cosmological simulations, concluding that major mergers are insufficient to explain the observed increase in ET population. \citet{lotz08ff} use morphological indices to study the major merger rate to $z \sim 1.2$ in a $B-$band selected sample ($L_{B} > 0.4L_{B}^{*}$) and compare it against the evolution of E/S0/Sa galaxies. They find that all the ET evolution can be explained by major mergers, although the effect of observational errors that tend to overestimate systematically the merger fraction by morphological indices \citep{clsj08ml} makes their result an upper limit.

Summarizing, our results are in good agreement with previous works when a similar mass selection is applied. At higher stellar masses the picture is different: the merger fraction depends on stellar mass \citep{deravel09,bundy09}, having $M_{\star} \gtrsim 10^{11}\ M_{\odot}$ galaxies higher pair fractions than less
massive ones, and being red pairs more common at these masses \citep{bundy09}. This suggests dry mergers as an important process in the evolution of massive
galaxies since $z \sim 1$ \citep[e.g.][]{bell04,lin08,ilbert10}. The size
\citep{trujillo07,buitrago08,vandokkum08,vanderwel08esize} and velocity
dispersion evolution \citep{cenarro09} of $M_{\star} \gtrsim 10^{11}\ M_{\odot}$
ET galaxies since $z \sim 2$ also support the importance of mergers, specially
the impact of minor mergers in the evolution of these systems
\citep{bezanson09,naab09,hopkins09size}. This problem was also analyzed by \citet{eliche09}, who modeled the evolution of luminosity function backwards in time for $M_{\star} \gtrsim 10^{11}\ M_{\odot}$ galaxies, selected according to their colors (red/blue/total) and their morphologies.  They find that the observed luminosity function evolution can be naturally
explained by the observed gas-rich and dry major merger rates, and that 50\%--60\% of
today's E/S0 in this mass range were formed by major mergers at $0.8 < z < 1$,
with a small number evolution since $z = 0.8$ (see also \citealt{cristobal09,ilbert10}). Note that the gas-rich major merger fractions assumed by \citet{eliche09} are those
from L09 for $B$-band selected galaxies ($M_B \leq -20$), which were obtained in
a similar way as the merger fractions used through this paper. 

This makes $M_{\star}^{*} \sim 10^{11}\ M_{\odot}$ \citep{pgon08} a transition mass at $z \lesssim 1$: at higher masses major mergers are an important process in the evolution of ET galaxies, while at lower masses other mechanisms dominate the observed evolution \citep[see also][]{drory08,vanderwel09,oesch09}.

\subsection{Which Processes are Responsible of Early-type Rise?}
Since gas-rich major mergers are not the dominant process in the late- to early-type transformation at intermediate masses, two possibilities remains: minor mergers and secular processes. We measure the structure in the $B$-band rest frame, so the increase in the S\'ersic
index that we observe can be either the result of an increase in the mass of the spheroidal
component of the galaxies, or due to the decline of the star formation in the disk of late types. In the former,
minor mergers increase the S\'ersic index of galaxies \citep{eliche06} and multiple minor mergers can lead to ETs \citep{bournaud07,hopkins09disk}, while
secular evolution, e.g. bars and disk instabilities, produces pseudobulges
\citep{kormendy04,fisher09,combes09}. In the latter, minor mergers can modify the gas distribution of the galaxy, shutting down the star formation \citep{bekki98},
whereas gas exhaustion or the prevention of star formation due to the
stabilization of the gas in a bulge-dominated galaxy (morphological quenching,
\citealt{martig09}) is an example of a secular process. To obtain new clues about ET evolution, we study the star formation properties of ET and LT galaxies in GOODS-S in a forthcoming paper (C. L\'opez-Sanjuan et al., in preparation).

\section{Conclusions}\label{conclusion}
We have studied the structure of $M_B \leq -20$ and $M_{\star} \geq
10^{10}\ M_{\odot}$ galaxies in GOODS-S at $z \lesssim 1$. We use the position
of galaxies in the concentration--asymmetry plane to segregate them in ET
galaxies (bulge dominated, E/S0/Sa), LT galaxies (disk dominated and
irregular, Sb-Irr), and gas-rich major mergers (see L09, for details). We find
that:
\begin{enumerate}
\item The ET fraction increases with cosmic time in both luminosity- and
mass-selected samples, while the LT fractions decrease;
\item The number density of LT $M_B \leq -20$ galaxies decreases with
cosmic time, while ET number density is roughly constant. We infer that
star
formation is located primarily in late spirals and irregulars, instead that in
major merger systems. These systems account for a $\sim 20$\%$-30$\% of the
total star formation rate in the range $0.4 \leq z < 1.1$, while the star
formation triggered directly by interactions in this range is $f^{\rm
tri}_{{\rm SF},{\rm m}} \sim 8$\%;

\item The number density of ET $M_{\star} \geq 10^{10}\ M_{\odot}$
galaxies increases by a factor of 5 since $z \sim 1$. This implies that we need a
structural transformation in the range $0 < z < 1$ between LT galaxies,
which form stars actively, and ET galaxies, in which stellar mass is
located;

\item When we compare the observed structural evolution with the gas-rich merger rate
in the range $0 < z < 1$, we obtain that only $\sim$20\% of the newly formed ET
galaxies that appear since $z = 1$ can be gas-rich major merger remnants, whereas
in the range $1 < z < 1.5$ these mergers can explain all the inferred
structural evolution. This suggests minor mergers and secular processes as the principal mechanisms in the rise of intermediate-mass ($M_{\star} \sim 4 \times 10^{10}\ M_{\odot}$) ET galaxies since $z \sim 1$, while gas-rich major mergers may be the
dominant process at higher redshifts ($z \gtrsim 1$) and masses ($M_{\star} \gtrsim 10^{11}\ M_{\odot}$).
\end{enumerate}

We study the star formation properties of GOODS-S galaxies in a forthcoming paper (C. L\'opez-Sanjuan et al., in
preparation) to obtain new clues about ETs rise since $z \sim 1$. In addition, the study of structure in red bands, more related to the stellar mass than the $B$ band used in this paper, are needed to
understand which phenomenon (bulge growth or disk fading) is responsible for the observed structural evolution.

\acknowledgments
We dedicate this paper to the memory of our six IAC colleagues and friends who
met with a fatal accident in Piedra de los Cochinos, Tenerife, in February 2007,
with a special thanks to Maurizio Panniello, whose teachings of \texttt{python}
were so important for this paper. 

We thank the anonymous referee for suggestions that improved the paper, and Ignacio Trujillo, Carmen Eliche-Moral, Rub\'en Sanchez-Janssen, Lilian Dom\'{\i}nguez-Palmero, and Mercedes Prieto for useful discussions.

This work was supported by the Spanish Programa Nacional de Astronom\'\i a y
Astrof\'{\i}sica through the project number AYA2006--12955, AYA2006--02358, and AYA
2006--15698-C02-02. This work was partially funded by the Spanish MEC under the
Consolider-Ingenio 2010 Program grant CSD2006-00070: First Science with the GTC
(http://www.iac.es/consolider-ingenio-gtc/).

This work is based on {\it HST}/ACS images from GOODS {\it HST} Treasury
Program, which is supported by NASA through grants HST-GO-09425.01-A and
HST-GO-09583.01, and in part on observations made with the {\it Spitzer Space
Telescope}, which is operated by the Jet Propulsion Laboratory, Caltech under
NASA contract 1407.

P. G. P. G. acknowledges support from the Ram\'on y Cajal Program financed by
the Spanish Government and the European Union.

{\it Facilities:} \facility{HST (ACS)}

\bibliography{apj-jour,biblio}
\bibliographystyle{apj}
\end{document}